\documentclass[ALICE,manyauthors]{cernphprep}
\pdfoutput=1
\usepackage{hyperref}
\usepackage{lineno}

\usepackage[english]{babel}
\usepackage{graphicx}
\usepackage{verbatim}
\usepackage{amsfonts}
\usepackage{makeidx}
\usepackage{color}
\usepackage{amsmath}
\usepackage{lineno}
\usepackage{epsfig}
\usepackage{epstopdf}
\usepackage{hyperref}
\usepackage{cite}
\usepackage{textcomp}
\newcommand{\sqrtsNN}{\sqrt{s_{\rm \scriptscriptstyle NN}}}

\newcommand{\GeV}{\mathrm{GeV}}
\newcommand{\TeV}{\mathrm{TeV}}

\newcommand{\mev}{\mathrm{MeV}}
\newcommand{\gev}{\mathrm{GeV}}

\newcommand{\tev}{\mathrm{TeV}}

\newcommand{\PbPb}{\mbox{Pb--Pb}}

\newcommand{\RAA}{R_{\rm AA}}

\newcommand{\pt}{p_{\rm T}}

\newcommand{\DtoKpi}{{\rm D}^0 \to {\rm K}^-\pi^+}
\newcommand{\DtoKpipi}{{\rm D}^+\to {\rm K}^-\pi^+\pi^+}
\newcommand{\DstartoDpi}{{\rm D}^{*+} \to {\rm D}^0 \pi^+}

\newcommand{\DstophipitoKKpi}{{\rm D_s^{+}\to \phi\pi^+\to K^-K^+\pi^+}}

\newcommand{\Dzero}{{\rm D^0}}

\newcommand{\Dstar}{{\rm D^{*+}}}

\newcommand{\Dplus}{{\rm D^+}}

\newcommand{\Ds}{{\rm D_s^+}}

\begin{document}%

\begin{titlepage}
\PHyear{2017}
\PHnumber{153}      
\PHdate{3 July}  
%

\title{D-meson azimuthal anisotropy in mid-central Pb--Pb collisions \\ at $\mathbf{\sqrtsNN=5.02}$~TeV}
\ShortTitle{D-meson azimuthal anisotropy in Pb--Pb collisions at $\sqrtsNN=5.02~\tev$}   

\Collaboration{ALICE Collaboration\thanks{See Appendix~\ref{app:collab} for the list of collaboration members}}
\ShortAuthor{ALICE Collaboration} 

\begin{abstract}
The azimuthal anisotropy coefficient $v_2$ of prompt $\Dzero$, $\Dplus$, $\Dstar$ and
$\Ds$ mesons was measured in mid-central (30--50\% centrality class) Pb--Pb collisions at a centre-of-mass energy per nucleon pair $\sqrtsNN = 5.02~\TeV$, with the ALICE detector at the LHC. 
The D mesons were reconstructed via their hadronic decays at mid-rapidity, $|y|<0.8$, in the transverse momentum interval $1<\pt<24~\GeV/c$. 
The measured D-meson $v_2$ has similar values as that of charged pions. 
The $\Ds$ $v_2$, measured for the first time, is found to be compatible with that of non-strange D mesons.
The measurements are compared with theoretical calculations of charm-quark 
transport in a hydrodynamically expanding medium and have the potential to constrain medium parameters.

\end{abstract}
\end{titlepage}
\setcounter{page}{2}

Quantum Chromodynamics predicts that strongly-interacting matter under extreme conditions of high temperature and energy density undergoes a transition from the hadronic phase to a color-deconfined medium, called Quark--Gluon Plasma (QGP)~\cite{Karsch:2006xs,Borsanyi:2010bp,Bazavov:2011nk,FLORKOWSKI2016669,PhysRevD.85.054503}. 
Heavy-ion collisions at ultra-relativistic energies provide suitable conditions for the QGP formation and for characterizing its properties. 

Heavy quarks (charm and beauty) are predominantly produced in hard scatterings before the QGP formation~\cite{BraunMunzinger:2007tn,Andronic:2015wma}. Therefore, they experience all stages of the medium evolution, interacting with its constituents via elastic~\cite{Braaten:1991we} and inelastic (radiation of gluons)~\cite{Gyulassy:1990ye,Baier:1996sk} processes (see~\cite{Andronic:2015wma,Prino:2016cni} for recent reviews).

Evidence of in-medium interactions and energy loss of charm quarks is provided by the strong modification of the transverse momentum ($\pt$) distributions of heavy-flavor hadrons in heavy-ion collisions with respect to pp collisions. 
A large suppression of heavy-flavor hadron yields was observed for $\pt> 4$--5$~\GeV/c$ in central nucleus--nucleus collisions at RHIC~\cite{Adare:2010de,Abelev:2006db,Adamczyk:2014uip,Adler:2005xv} and LHC~\cite{Adam:2015sza,Abelev:2012qh,Adam:2016khe,Adam:2016wyz,Khachatryan:2016ypw}. 

Measurements of anisotropies in the azimuthal distribution of heavy-flavor hadrons assess the transport properties of the medium.
The collective dynamics of the expanding medium converts the initial-state spatial anisotropy~\cite{Qin:2010pf}
into final-state particle momentum anisotropy. This anisotropy is characterized by the Fourier coefficients $v_n$ of the distribution of the particle azimuthal angle $\varphi$ relative to the initial-state symmetry plane angle $\Psi_n$ (for the $n^{\rm th}$ harmonic)~\cite{Voloshin:1994mz,Poskanzer:1998yz}.
In non-central collisions, the largest contribution corresponds to $v_2 = \langle {\rm cos}[2(\varphi-\Psi_2)]\rangle$, called elliptic flow~\cite{Ollitrault:1992bk,Poskanzer:1998yz}. 
The D-meson $v_2$ at low $\pt$ provides insight into the possible collective flow imparted by the medium to charm quarks~\cite{Batsouli:2002qf}, while at high $\pt$ it is sensitive to the path-length dependence of parton energy loss~\cite{Gyulassy:2000gk,Shuryak:2001me}. 
At low and intermediate $\pt$, a fraction of charm quarks could hadronize via recombination with light quarks from the medium, leading to an increase of the D-meson $v_2$ with respect to that of charm quarks~\cite{Molnar:2004ph,Andronic:2003zv,Greco:2003vf}; the comparison of the $v_2$ of D mesons without and with strange-quark content could be sensitive to these effects and to the charm coupling to the QGP and hadronic matter~\cite{He:2012df}. 

A positive heavy-flavor elliptic flow was observed in Au--Au collisions at $\sqrtsNN = 200~\GeV$~\cite{Adare:2010de,Adamczyk:2014yew,Adamczyk:2017xur}
and in Pb--Pb collisions at $\sqrtsNN=2.76$ $\TeV$~\cite{Abelev:2013lca,Abelev:2014ipa,Adam:2016ssk,Adam:2015pga,Khachatryan:2016ypw}.
Calculations based on heavy-quark transport in a hydrodynamically-expanding medium describe the measurements~\cite{Uphoff:2012gb,He:2014cla,Monteno:2011gq,Djordjevic:2015hra,Cao:2013ita,Song:2015ykw,Nahrgang:2013xaa,Uphoff:2014hza,Beraudo:2014boa,Cao:2017hhk}. 
Precise measurements of heavy-flavor $v_2$ constrain model parameters, e.g.\,the heavy-quark spatial diffusion coefficient $D_s$ in the QGP, which is related to the relaxation (equilibration) time of heavy quarks $\tau_{\rm Q} = \frac{m_{\rm Q}}{T}D_s$, where $m_{\rm Q}$ is the quark mass and $T$ is the medium temperature~\cite{Moore:2004tg}.

In this Letter, we report on the $v_2$ of $\Dzero$, $\Dplus$, $\Dstar$ and, for the first time at the LHC, of $\Ds$ mesons, and their antiparticles, in Pb--Pb collisions at $\sqrtsNN = 5.02$~$\TeV$, for the 30--50\% centrality class.
The analysis uses $\PbPb$ collisions collected with the ALICE detector~\cite{Aamodt:2008zz, Abelev:2014ffa} in 2015. The interaction trigger consisted in coincident signals in the two scintillator arrays of the V0 detector, covering full azimuth in the pseudorapidity ($\eta$) regions $-3.7 < \eta < -1.7$ and $2.8 < \eta < 5.1$. Events from beam--gas interactions are removed using time information from the V0 and the neutron Zero-Degree Calorimeters. Only the events with a primary vertex reconstructed within $\pm 10$~cm from the detector centre along the beam direction are analysed. Events are selected in the centrality class 30--50\%, defined in terms of percentiles of the hadronic Pb--Pb cross section, using the amplitude of the V0 signals~\cite{Abelev:2013qoq, Adam:2015ptt}. The number of selected events is $20.7 \times 10^6$, corresponding to an integrated luminosity $L_{\rm int} \approx 13 \mu{\rm b}^{-1}$~\cite{Adam:2015ptt}.

The D mesons and their antiparticles are reconstructed using the decay channels $\DtoKpi$, $\DtoKpipi$, $\DstartoDpi$, and $\DstophipitoKKpi$.
The analysis procedure~\cite{Abelev:2014ipa, Adam:2015jda} searches for decay vertices displaced from 
the interaction vertex, exploiting the mean proper decay lengths of about 123, 
312 and 150~$\mu{\rm m}$ of $\Dzero$, $\Dplus$ and $\Ds$ 
mesons~\cite{Olive:2016xmw}.
Charged-particle tracks are reconstructed using the Inner Tracking System (ITS) and the Time Projection Chamber (TPC), which 
are located within a solenoid magnet that provides a 0.5~T field, parallel to the beam direction. 
$\Dzero$, $\Dplus$ and $\Ds$ candidates are defined using pairs and triplets of tracks with $|\eta| < 0.8$, $\pt
> 0.4~\GeV/c$, 70--159 TPC space points and 2--6 hits in the ITS (at least one in the two innermost layers). $\Dstar$ candidates are formed by combining $\Dzero$ candidates with tracks with $|\eta| < 0.8$, $\pt > 0.1~\GeV/c$ and at least three ITS hits. The selection of tracks with $|\eta|<0.8$ limits the D-meson acceptance in rapidity, which varies from $|y|<0.6$ for $\pt = 1~\GeV/c$ to $|y|<0.8$ for $\pt > 5~\GeV/c$. The main variables used to select the D candidates are the separation between the primary and decay vertices, the displacement of the tracks from the primary vertex and the pointing of the reconstructed D-meson momentum to the primary vertex.
For the selection of $\DstophipitoKKpi$ decays, one of the two pairs of opposite-sign tracks must have an invariant mass compatible with the $\phi$-meson mass~\cite{Olive:2016xmw}. 
Further background reduction results from the particle identification. A $\pm 3\,\sigma$ window around the expected mean values of the specific ionisation energy loss ${\rm d}E/{\rm d} x$ in the TPC gas and time-of-flight from the interaction point to the Time-Of-Flight (TOF) detector is used for each track, where $\sigma$ is the resolution on the two variables.
For $\Ds$ candidates, tracks not matched to a hit in the TOF (mostly at low momentum) are required to have a $2\,\sigma$ compatibility with the expected ${\rm d}E/{\rm d} x$ in the TPC. 
These selections result in signal-to-background ratios between 0.04 and 2.8 and a statistical
significance between 3 and 20, depending on the D-meson species and $\pt$.

The second harmonic symmetry plane ${\rm \Psi}_2$ is estimated, for each collision, 
by the Event Plane (EP) angle, denoted ${\psi}_2$, using the signals produced 
by charged particles in the eight azimuthal sectors of each V0 array. Effects of non-uniform V0
acceptance are corrected for using the gain equalisation method~\cite{PhysRevC.77.034904}.
The $v_{\rm 2}$ was calculated by classifying D mesons 
in two groups, according to their azimuthal angle relative to the EP $\Delta \varphi=\varphi_{\rm D} - \psi_{2}$: 
in-plane ($] -\frac{\pi}{4},\frac{\pi}{4}]$ and $]\frac{3\pi}{4},\frac{5\pi}{4}]$) and out-of-plane 
($] \frac{\pi}{4},\frac{3\pi}{4}]$ and $]\frac{5\pi}{4},\frac{7\pi}{4}]$).
Integrating the d$N/$d$\varphi$ distribution in these two  $\Delta \varphi$ intervals, $v_{\rm 2}$
can be expressed as~\cite{Abelev:2014ipa}:
\begin{equation} 
\label{eq:twobins} 
v_2\{{\rm EP}\} =\frac{1}{R_2}\frac{\pi}{4}\frac{N_{\textnormal{in-plane}}-N_{\textnormal{out-of-plane}}}{N_{\textnormal{in-plane}}+N_{\textnormal{out-of-plane}}}\ ,
\end{equation}
where $N_{\textnormal{in-plane}}$ and $N_{\textnormal{out-of-plane}}$ are the D-meson yields in the two $\Delta \varphi$ intervals.  
The factor $\frac{1}{R_2}$ is the correction for the resolution in the
estimation of the symmetry plane $\Psi_2$ via the EP angle $\psi_2$.
It is calculated using three sub-events of charged particles in the V0 and in the positive and negative $\eta$ regions of the TPC~\cite{Poskanzer:1998yz}.
The separation of at least 0.9 units of pseudorapidity ($|\Delta\eta|>0.9$) between the D mesons 
and the particles used in the $\psi_2$ calculation
suppresses non-flow contributions to $v_2$ (i.e.\,correlations not induced by the collective expansion but rather by decays and jet production).

Simulations showed that the D-meson reconstruction and selection efficiencies do not depend on $\Delta\varphi$~\cite{Abelev:2014ipa}, 
therefore Eq.~(\ref{eq:twobins}) can be applied using the D-meson raw yields, without an efficiency correction.
The raw yields were obtained from fits to the $\Dzero$, 
$\Dplus$ and $\Ds$ candidate invariant-mass distributions and to the mass difference 
$\Delta M = M (\mathrm{K} \pi \pi) - M(\mathrm{K} \pi)$ distributions for $\Dstar$ candidates. 
In the fit function, the signal was modelled with a Gaussian and the background with an 
exponential term for $\Dzero$, $\Dplus$ and $\Ds$ candidates and with the function 
$a \sqrt{\Delta M - m_{\pi}} \cdot {\rm e}^{b(\Delta M - m_{\pi})}$ for $\Dstar$ candidates. 
The mean and the width of the Gaussian were fixed to those obtained from a fit to
the sum of the invariant-mass distributions in the two $\Delta \varphi$ intervals, where the signal has higher statistical significance. 
In the $\Dzero$ invariant-mass fit, the contribution of signal candidates with the wrong K--$\pi$ mass assignment 
(about 2--5\% of the raw signal depending on $\pt$) was taken into account by including an additional term, 
parametrised from simulations with a double-Gaussian shape, in the fit function~\cite{Abelev:2014ipa}.

The measured D-meson yield includes the contributions of prompt D mesons, from c-quark hadronization or strong decays 
of $\rm D^{*}$  states, and of
feed-down D mesons from beauty-hadron decays. 
The observed $v_{\rm 2}$, measured with Eq.~(\ref{eq:twobins}), is a linear combination of the prompt and feed-down contributions:
$v_{\rm 2}^{\rm obs} = f_{\rm prompt}\cdot v_{\rm 2}^{\rm prompt} + (1-f_{\rm prompt})v_{\rm 2}^{\rm feed\textnormal{-}down}$,
where $f_{\rm prompt}$ is the fraction of prompt D mesons in the raw yields and
$v_{\rm 2}^{\rm feed\textnormal{-}down}$
is the elliptic flow of D mesons from beauty-hadron decays.
To calculate $v_{\rm 2}^{\rm prompt}$, a hypothesis on $v_{\rm 2}^{\rm feed\textnormal{-}down}$ is used.
The measured $v_2$ of non-prompt J/$\psi$~\cite{Khachatryan:2016ypw} and the available model calculations~\cite{Aichelin:2012ww,Uphoff:2012gb,Greco:2007sz} 
suggest that $0<v_{\rm 2}^{\rm feed\textnormal{-}down}<v_{\rm 2}^{\rm prompt}$.
Assuming a uniform probability distribution of $v_{\rm 2}^{\rm feed\textnormal{-}down}$ in this interval,
the central value for $v_{\rm 2}^{\rm prompt}$ is calculated considering $v_{\rm 2}^{\rm feed\textnormal{-}down}=v_{\rm 2}^{\rm prompt}/2$, 
thus $v_{\rm 2}^{\rm prompt}= 2\,v_{\rm 2}^{\rm obs}/(1+f_{\rm prompt})$.
The $f_{\rm prompt}$ fraction is estimated, as a function of $\pt$, as described in~\cite{ALICE:2012ab}, 
using the FONLL~\cite{Cacciari:2012ny} calculation for the beauty-hadron cross section, 
the beauty-hadron decay kinematics from EvtGen~\cite{Lange:2001uf}, 
the reconstruction efficiencies for feed-down D mesons from simulation, and a hypothesis for the nuclear modification factor of the feed-down D mesons, $R_{\rm AA}^{\rm feed\textnormal{-}down}$.
The nuclear modification factor is defined as the ratio of the 
$p_{\rm T}$-differential yields in nucleus--nucleus and pp collisions scaled 
by the average number of nucleon--nucleon collisions in the considered 
centrality class~\cite{Miller:2007ri}.
By comparison of the $R_{\rm AA}$ of prompt D mesons~\cite{Adam:2015nna} and
J/$\psi$ mesons from beauty-hadron decays~\cite{Khachatryan:2016ypw} in Pb--Pb collisions at $\sqrtsNN = 2.76~\TeV$, 
the assumptions $R_{\rm AA}^{\rm feed\textnormal{-}down}=2\,R_{\rm AA}^{\rm prompt}$ for non-strange D mesons
and $R_{\rm AA}^{\rm feed\textnormal{-}down}=\,R_{\rm AA}^{\rm prompt}$ for the $\Ds$ meson are made to compute $f_{\rm prompt}$.

The systematic uncertainty from feed-down on $v_{2}^{\rm prompt}$ was estimated by varying the central value of 
$v_{2}^{\rm feed\textnormal{-}down}=v_{2}^{\rm prompt}/2$ by $ \pm v_{2}^{\rm prompt}/\sqrt{12}$, corresponding to 
$\rm \pm 1\,RMS$ of a uniform distribution in $(0,\,v_{2}^{\rm prompt})$.
The uncertainty on $f_{\rm prompt}$ was obtained from the
variation of the FONLL calculation parameters, and from the variation of the 
$R_{\rm AA}^{\rm feed\textnormal{-}down}$ hypothesis in  
\mbox{ $1< R_{\rm AA}^{\rm feed\textnormal{-}down}/R_{\rm AA}^{\rm prompt}<3$} 
for non-strange D mesons~\cite{Adam:2015sza} and  
$\frac{1}{3}< R_{\rm AA}^{\rm feed\textnormal{-}down}/R_{\rm AA}^{\rm prompt}<3$ for $\Ds$ mesons~\cite{Adam:2015jda}. 
The value of the absolute systematic uncertainty from feed-down ranges from 0.001 to 0.030.

The other sources of systematic uncertainty are related to
the signal extraction from the invariant-mass distribution, non-flow effects, and centrality dependence in the EP resolution correction $R_2$.

The signal extraction uncertainty was estimated by varying the background fit function and leaving the Gaussian width and mean as free parameters in the fit.
Furthermore, an alternative method for the yield extraction based on counting the histogram entries in the signal invariant-mass region,
after subtracting the background estimated from a fit to the side bands, was considered.
The absolute systematic uncertainties on $v_2$ due to the yield extraction range from 
0.005 to 0.040 for $\Dzero$, $\Dplus$ and $\Dstar$, and from 0.015 to 0.070 for $\Ds$ mesons.
As a check of a possible efficiency dependence on $\Delta \varphi$, 
the analysis was repeated
with different selection criteria and no systematic effect was observed.

The EP resolution correction $R_2$ depends on collision centrality~\cite{Abelev:2014ipa}.
The value used in Eq.~(\ref{eq:twobins}) was computed assuming a uniform distribution of the D-meson yield within the centrality class.
This value was compared with those obtained from the weighted averages of the $R_2$ values in narrow centrality 
intervals, using as weights either the D-meson yields or the number of 
nucleon--nucleon collisions. 
In addition,  to account for the presence of possible non-flow effects in the estimation of $R_2$, 
its value was re-computed using two different pseudorapidity gaps between the
sub-events of the TPC tracks with positive/negative $\eta$.  A systematic uncertainty of 2\% on $R_2$ was estimated.

\begin{figure*}[!t]
\begin{center}
\includegraphics[width=.65\textwidth]{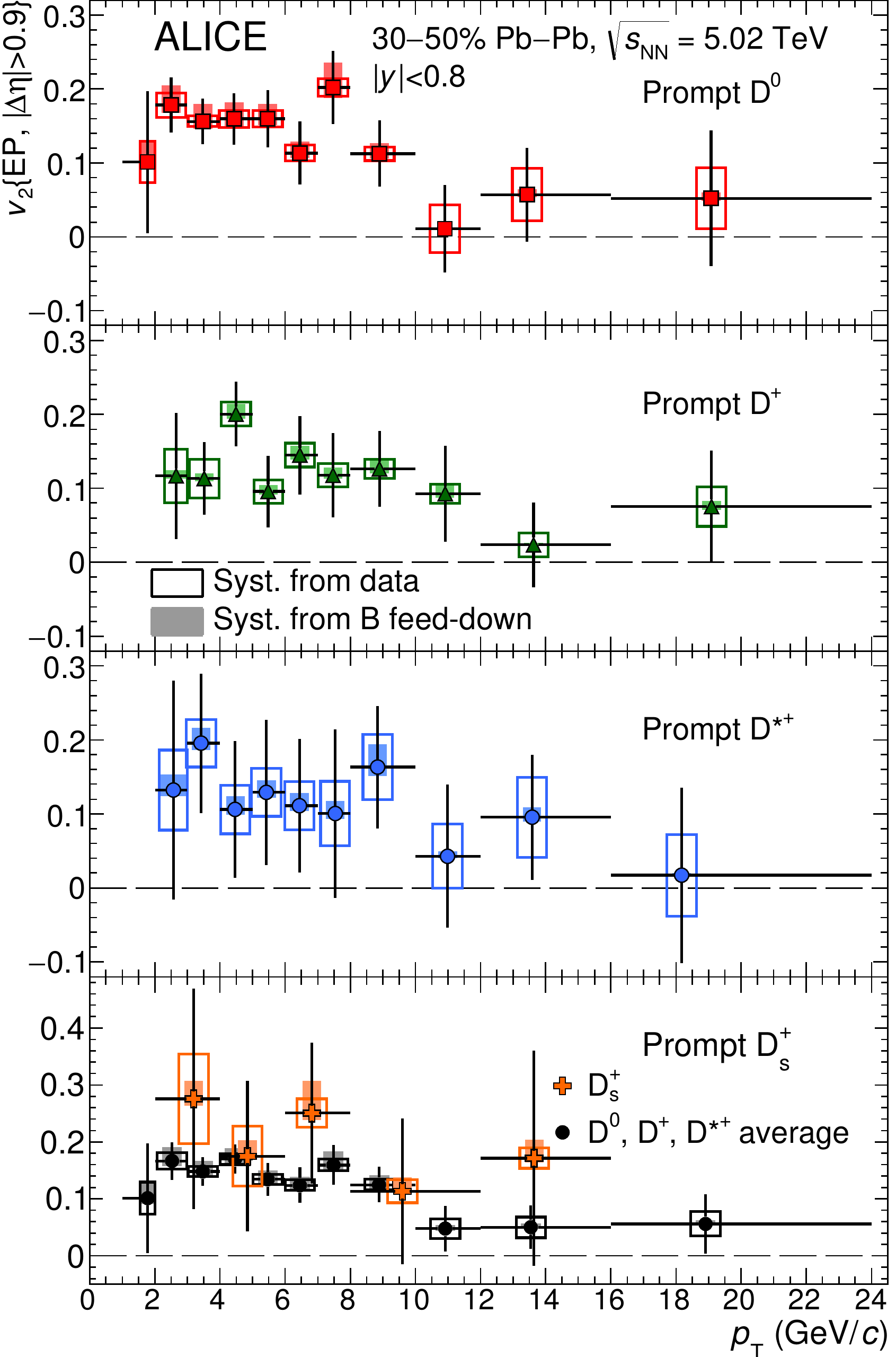}
\caption{Elliptic flow coefficient as a function of $\pt$ for prompt $\Dzero$, $\Dplus$, 
$\Dstar$ and $\Ds$ mesons and their charge conjugates for $\PbPb$ collisions in the centrality class 30--50\%.
The bottom panel also shows the average
$v_2$ of $\Dzero$, $\Dplus$ and $\Dstar$. 
Vertical bars represent the statistical uncertainty, empty boxes the systematic 
uncertainty associated with the D-meson anisotropy measurement and the event-plane 
resolution. Shaded boxes show the feed-down uncertainty.}
\label{fig:v2_4mesons} 
\end{center}
\end{figure*}

\begin{figure}[!t]
\begin{center}
\includegraphics[width=0.65\textwidth]{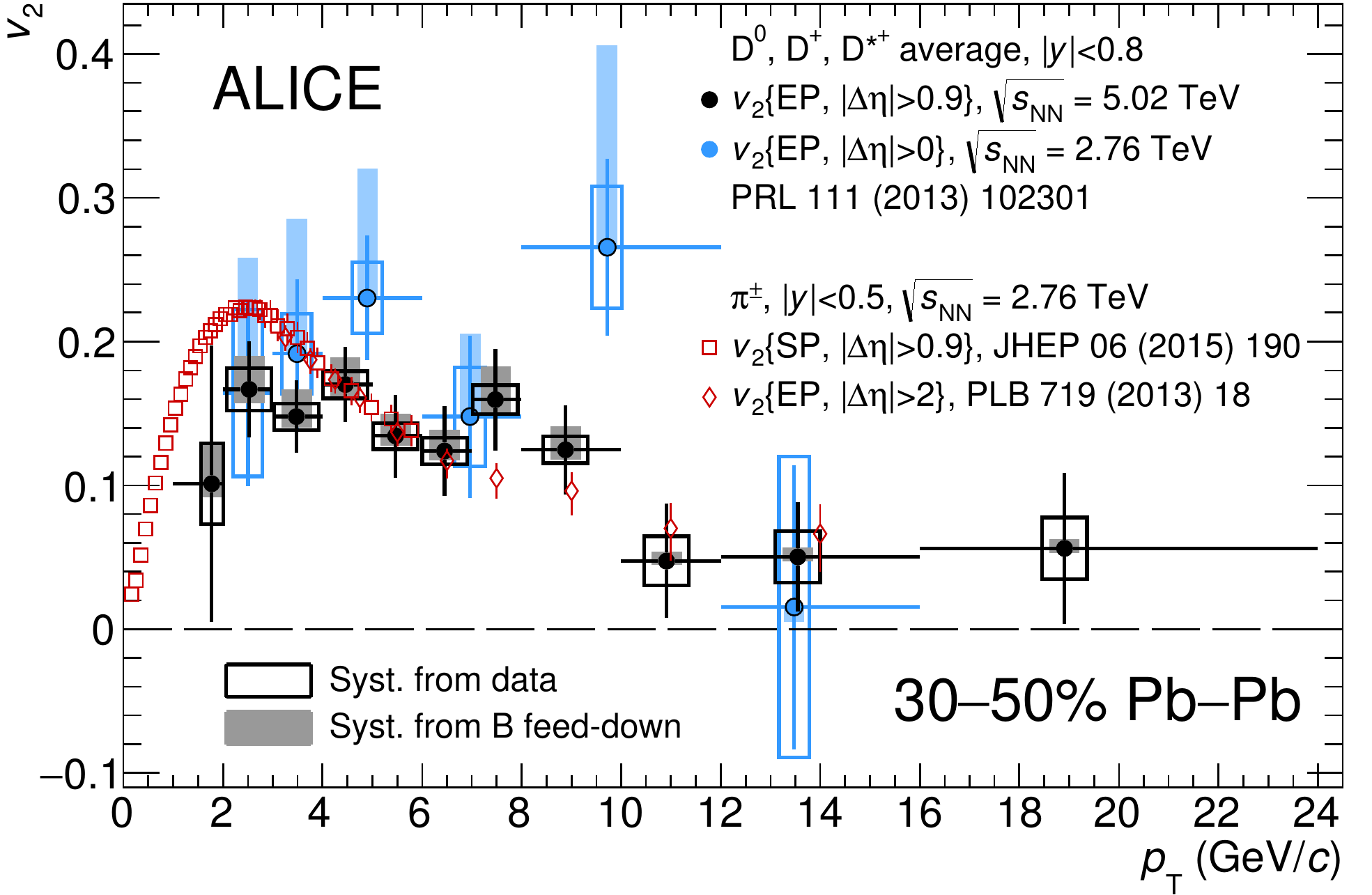}
\caption{Average of $\Dzero$, $\Dplus$ and $\Dstar$ $v_2$ as a function of 
$\pt$ at $\sqrtsNN=5.02~\tev$, compared with the same measurement at $\sqrtsNN=2.76~\tev$~\cite{Abelev:2013lca} and to the $\pi^\pm$ $v_2$ measured with the EP method~\cite{Abelev:2012di,ALICE-PUBLIC-2015-003} and with the scalar production (SP) method~\cite{Abelev:2014pua}.}
\label{fig:v2withpions} 
\end{center}
\end{figure}

The $v_2$ of prompt $\Dzero$, $\Dplus$, $\Dstar$ and $\Ds$ mesons in
the 30--50\% centrality class is shown in Fig.~\ref{fig:v2_4mesons}.
The symbols are positioned at the average $\pt$ of the 
reconstructed D mesons: this value was determined as the average of the $\pt$ distribution of candidates in the signal invariant-mass region, 
after subtracting the contribution of the background candidates estimated from the side bands.
The $v_2$ of $\Dzero$, $\Dplus$ and $\Dstar$ are consistent and they are larger than zero in $2<\pt<10~\gev/c$. The $\Dzero$ $v_2$ is 
compatible with the measurement by the CMS collaboration~\cite{Sirunyan:2017plt}.
The average of the $v_2$ measurements for $\Ds$ mesons in the three $\pt$ intervals within $2<\pt<8~\gev/c$ is positive with a significance of 2.6\,$\sigma$,
where $\sigma$ is the uncertainty of the average $v_2$, calculated using quadratic propagation for the statistical and uncorrelated systematic uncertainties 
(signal extraction) and linear propagation for the correlated systematic uncertainties ($R_2$ and feed-down correction).
The average $v_2$ and $\pt$ of $\Dzero$, $\Dplus$ and $\Dstar$, shown in the bottom panel of Fig.~\ref{fig:v2_4mesons}, was 
computed using the inverse of the squared statistical uncertainties as weights. 
The systematic uncertainties were propagated
treating the $R_2$  and feed-down contributions as correlated among D-meson species. 

Fig.~\ref{fig:v2withpions} shows that
the average $v_2$ of $\Dzero$, $\Dplus$ and $\Dstar$ at $\sqrtsNN=5.02~\tev$ is compatible 
with the same measurement at $\sqrtsNN=2.76~\tev$ ($L_{\rm int} \approx 6$~\textmu$\rm b^{-1}$)~\cite{Abelev:2013lca}, which has uncertainties
larger by a factor of about two compared to the new result at
$5.02~\tev$. Note that  the vertexing and tracking performance
improved in 2015 and in~\cite{Abelev:2013lca} the correction for feed-down was made with the assumption $v_2^\textrm{feed-down}=v_2^{\rm prompt}$. The assumption 
used in the present analysis, $v_2^\textrm{feed-down}=v_2^{\rm prompt}/2$, would increase the 
values at $\sqrtsNN=2.76~\tev$ by about 10\%.

The average D-meson $v_2$ is also compared with the
$\pi^\pm$ $v_2$ at 
$\sqrtsNN=2.76~\tev$
measured with the EP method~\cite{Abelev:2012di,ALICE-PUBLIC-2015-003} considering a pseudorapidity separation of 2 units between  $\pi^\pm$ and the particles used to measure the EP angle, and the scalar-product method~\cite{Abelev:2014pua}, also based on 2-particle correlations.
The comparison of the D-meson $v_2$ at $\sqrtsNN=5.02~\tev$ and of the pion $v_2$ at $\sqrtsNN=2.76~\tev$ is justified 
by the observation that the $\pt$-differential $v_2$ of charged particles, which is dominated by the pion component, is 
compatible at these two energies~\cite{Adam:2016izf}.
The D-meson $v_2$ is similar to that of $\pi^\pm$ in the common $\pt$ interval (1--16~$\gev/c$)
and it is lower in the interval below $4~\gev/c$, the difference reaching about 2\,$\sigma$ in 2--4~$\gev/c$, where a mass ordering of $v_2$ is
observed  for light-flavor hadrons and described by hydrodynamical calculations~\cite{Abelev:2014pua}.

\begin{figure}[!t]
\begin{center}
\includegraphics[width=0.65\textwidth]{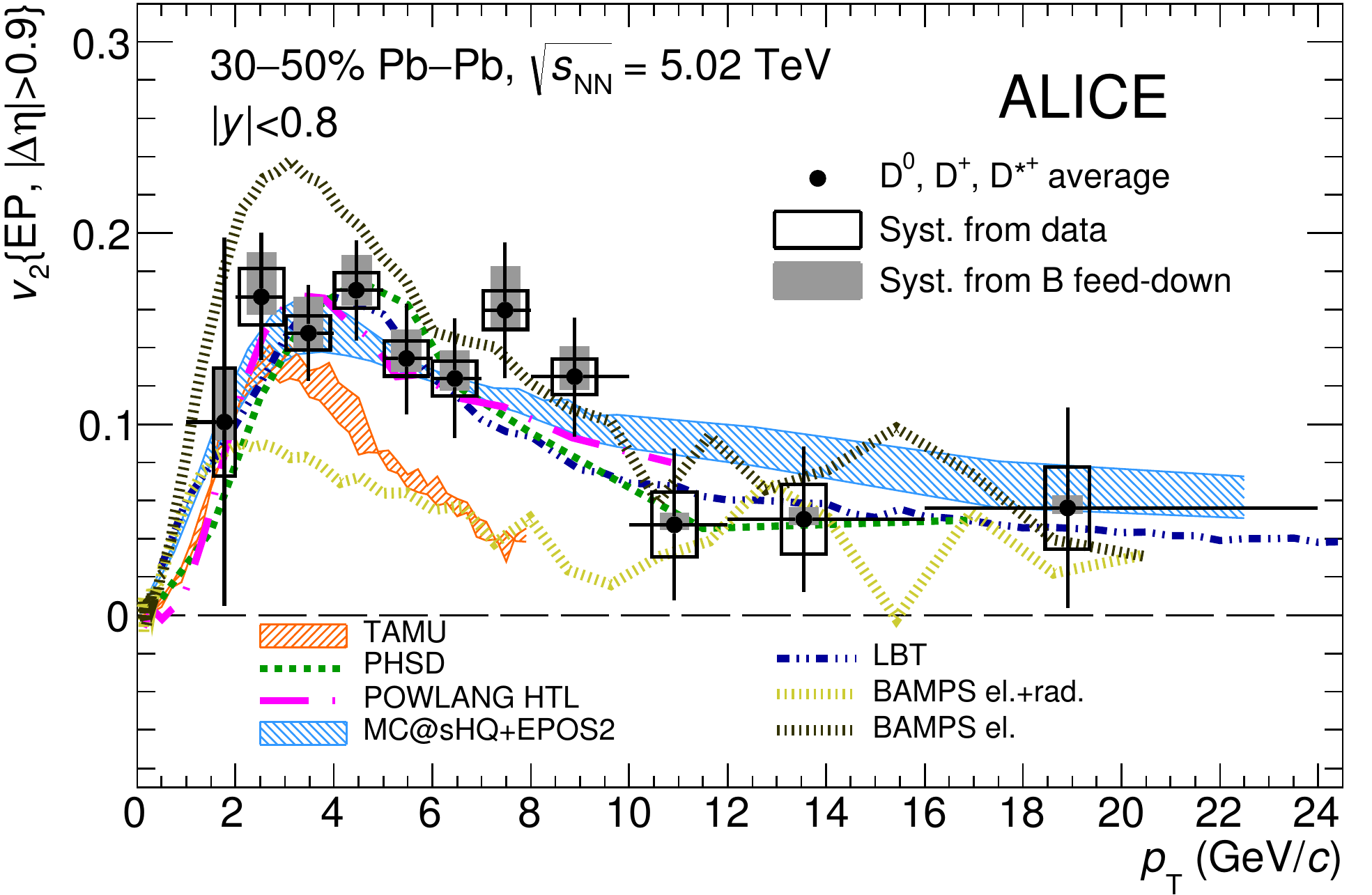}
\caption{Average of $\Dzero$, $\Dplus$ and $\Dstar$ $v_2$ as a function of 
$\pt$, compared with model
calculations~\cite{Uphoff:2014hza,Beraudo:2014boa,He:2014cla,Cao:2017hhk,Nahrgang:2013xaa,Song:2015ykw}. }
\label{fig:v2withmodels} 
\end{center}
\end{figure}

In Fig.~\ref{fig:v2withmodels}, the average $v_2$ of the three
non-strange D-meson species is compared with theoretical calculations
that include a hydrodynamical model for the QGP expansion (models that
lack this expansion 
underestimated the D-meson $v_2$ measurements at $\sqrtsNN=2.76~\tev$
in $2<\pt<6~\gev/c$~\cite{Abelev:2014ipa}). 
The BAMPS-el~\cite{Uphoff:2014hza},  POWLANG~\cite{Beraudo:2014boa}  and TAMU~\cite{He:2014cla} calculations include only collisional (i.e.\,elastic) interaction processes, while the BAMPS-el+rad~\cite{Uphoff:2014hza},  LBT~\cite{Cao:2017hhk}, MC@sHQ~\cite{Nahrgang:2013xaa} and PHSD~\cite{Song:2015ykw} calculations also include energy loss via gluon radiation. All calculations, with the exception of BAMPS, include hadronization via quark recombination, in addition to independent fragmentation. 
The MC@sHQ and TAMU results are displayed with their theoretical uncertainty band.
All calculations provide a fair description of the 
nuclear modification factor of D mesons in central Pb--Pb collisions at $\sqrtsNN=2.76~\tev$ in $1<\pt<8~\gev/c$~\cite{Adam:2015sza}.

The $v_2$ measurement at $\sqrtsNN=5.02~\tev$ is described by most of these calculations, in which the interactions with the 
hydrodynamically-expanding medium impart a positive $v_2$ to charm quarks.
The model-to-data consistency was quantified using the
reduced $\chi^2$ in the $\pt$ interval where all calculations are
available (2--8~$\gev/c$): the LBT, MC@sHQ, PHSD and POWLANG models
have $\chi^2/{\rm ndf}<1$, the TAMU, BAMPS-el+rad and BAMPS-el
models have a $\chi^2$/ndf of 4.1, 6.7 and 1.9,
respectively. The $\chi^2$ calculation includes the data uncertainties
and the model uncertainties when available. For BAMPS-el+rad, 
the low value of $v_2$ is caused by the absence
of the recombination contribution~\cite{Uphoff:2014hza}. 
For TAMU, the rapid decrease of $v_2$ with increasing $\pt$
is due to the lack of radiative energy loss, which is also
reflected in $\RAA$ values larger than the measured ones at high 
$\pt$~\cite{Adam:2015sza}. 
For most of these calculations, the medium effect on heavy quarks can
be expressed using the dimensionless quantity
$2\pi\,T\,D_s(T)$~\cite{Moore:2004tg}.  In the interval from the critical temperature for QGP formation $T_{\rm c}\approx 155~\mev$~\cite{Borsanyi:2010bp} to 2\,$T_{\rm c}$, the ranges of $2\pi\,T\,D_s(T)$ are: 1--2 for BAMPS-el, 6--10 for BAMPS-el+rad, 2--6 for LBT~\cite{Xu:2017hgt}, 1.5--4.5 for MC@sHQ~\cite{Andronic:2015wma}, 4--9 for PHSD~\cite{Song:2015ykw}, 7--18 for POWLANG~\cite{Prino:2016cni} and 4--10 for  TAMU~\cite{Andronic:2015wma}.  
The calculations that describe the data with $\chi^2/{\rm ndf} < 1$ use $2\pi\,T\,D_s(T)$ in the range of 1.5--7 at $T_{\rm c}$. Remarkably, this range is consistent with that obtained by the comparison of the $\Dzero$ $v_2$ in Au--Au collisions at $\sqrtsNN=200~\gev$ to model calculations~\cite{Adamczyk:2017xur}, and it includes the values obtained by lattice QCD calculations~\cite{Ding:2012sp,Banerjee:2011ra} which are independent of the collision energy because they encode a property of the medium evaluated at a fixed temperature.
The corresponding thermalisation time~\cite{Moore:2004tg} for charm
quarks is $\tau_{\rm charm}=\frac{m_{\rm charm}}{T}  D_s(T) \approx 3$--14~fm/$c$ with $T=T_{\rm c}$ and $m_{\rm charm}=1.5~\gev/c^2$. 
These values are comparable to the estimated decoupling time of the high-density system~\cite{Aamodt:2011mr}.
It should also be pointed out that the models differ in several aspects, related to the medium expansion and the heavy quark--medium interactions both in the QGP and in the hadronic phase.

In summary, we have presented a measurement of the elliptic 
flow $v_2$ of prompt $\Dzero$, $\Dplus$, $\Dstar$ and $\Ds$ mesons
in Pb--Pb collisions at $\sqrtsNN=5.02~\tev$.
The average $v_2$ of non-strange D mesons was measured with statistical and systematic uncertainties 
smaller by a factor about two with respect to our measurement at
$\sqrtsNN=2.76~\tev$. The results at the two energies are compatible within statistical
uncertainties. The $\Ds$ $v_2$ was for the first time measured at the LHC, 
although with a limited precision, and found to be compatible with that of non-strange D mesons.
The comparison of the D-meson $v_2$ with that of pions and with model calculations
indicates that low-momentum charm quarks take part in the collective motion of the 
QGP and that collisional interaction processes as well as recombination of charm and light quarks both contribute to the observed elliptic flow.
The calculations that describe the
measurements use heavy-quark spatial diffusion
coefficients in the range of $2\pi\,T\,D_s(T)\approx 1.5$--7 
at the critical temperature $T_{\rm c}$.

\newenvironment{acknowledgement}{\relax}{\relax}
\begin{acknowledgement}
\section*{Acknowledgements}

The ALICE Collaboration would like to thank all its engineers and technicians for their invaluable contributions to the construction of the experiment and the CERN accelerator teams for the outstanding performance of the LHC complex.
The ALICE Collaboration gratefully acknowledges the resources and support provided by all Grid centres and the Worldwide LHC Computing Grid (WLCG) collaboration.
The ALICE Collaboration acknowledges the following funding agencies for their support in building and running the ALICE detector:
A. I. Alikhanyan National Science Laboratory (Yerevan Physics Institute) Foundation (ANSL), State Committee of Science and World Federation of Scientists (WFS), Armenia;
Austrian Academy of Sciences and Nationalstiftung f\"{u}r Forschung, Technologie und Entwicklung, Austria;
Ministry of Communications and High Technologies, National Nuclear Research Center, Azerbaijan;
Conselho Nacional de Desenvolvimento Cient\'{\i}fico e Tecnol\'{o}gico (CNPq), Universidade Federal do Rio Grande do Sul (UFRGS), Financiadora de Estudos e Projetos (Finep) and Funda\c{c}\~{a}o de Amparo \`{a} Pesquisa do Estado de S\~{a}o Paulo (FAPESP), Brazil;
Ministry of Science \& Technology of China (MSTC), National Natural Science Foundation of China (NSFC) and Ministry of Education of China (MOEC) , China;
Ministry of Science, Education and Sport and Croatian Science Foundation, Croatia;
Ministry of Education, Youth and Sports of the Czech Republic, Czech Republic;
The Danish Council for Independent Research | Natural Sciences, the Carlsberg Foundation and Danish National Research Foundation (DNRF), Denmark;
Helsinki Institute of Physics (HIP), Finland;
Commissariat \`{a} l'Energie Atomique (CEA) and Institut National de Physique Nucl\'{e}aire et de Physique des Particules (IN2P3) and Centre National de la Recherche Scientifique (CNRS), France;
Bundesministerium f\"{u}r Bildung, Wissenschaft, Forschung und Technologie (BMBF) and GSI Helmholtzzentrum f\"{u}r Schwerionenforschung GmbH, Germany;
General Secretariat for Research and Technology, Ministry of Education, Research and Religions, Greece;
National Research, Development and Innovation Office, Hungary;
Department of Atomic Energy Government of India (DAE) and Council of Scientific and Industrial Research (CSIR), New Delhi, India;
Indonesian Institute of Science, Indonesia;
Centro Fermi - Museo Storico della Fisica e Centro Studi e Ricerche Enrico Fermi and Istituto Nazionale di Fisica Nucleare (INFN), Italy;
Institute for Innovative Science and Technology , Nagasaki Institute of Applied Science (IIST), Japan Society for the Promotion of Science (JSPS) KAKENHI and Japanese Ministry of Education, Culture, Sports, Science and Technology (MEXT), Japan;
Consejo Nacional de Ciencia (CONACYT) y Tecnolog\'{i}a, through Fondo de Cooperaci\'{o}n Internacional en Ciencia y Tecnolog\'{i}a (FONCICYT) and Direcci\'{o}n General de Asuntos del Personal Academico (DGAPA), Mexico;
Nederlandse Organisatie voor Wetenschappelijk Onderzoek (NWO), Netherlands;
The Research Council of Norway, Norway;
Commission on Science and Technology for Sustainable Development in the South (COMSATS), Pakistan;
Pontificia Universidad Cat\'{o}lica del Per\'{u}, Peru;
Ministry of Science and Higher Education and National Science Centre, Poland;
Korea Institute of Science and Technology Information and National Research Foundation of Korea (NRF), Republic of Korea;
Ministry of Education and Scientific Research, Institute of Atomic Physics and Romanian National Agency for Science, Technology and Innovation, Romania;
Joint Institute for Nuclear Research (JINR), Ministry of Education and Science of the Russian Federation and National Research Centre Kurchatov Institute, Russia;
Ministry of Education, Science, Research and Sport of the Slovak Republic, Slovakia;
National Research Foundation of South Africa, South Africa;
Centro de Aplicaciones Tecnol\'{o}gicas y Desarrollo Nuclear (CEADEN), Cubaenerg\'{\i}a, Cuba, Ministerio de Ciencia e Innovacion and Centro de Investigaciones Energ\'{e}ticas, Medioambientales y Tecnol\'{o}gicas (CIEMAT), Spain;
Swedish Research Council (VR) and Knut \& Alice Wallenberg Foundation (KAW), Sweden;
European Organization for Nuclear Research, Switzerland;
National Science and Technology Development Agency (NSDTA), Suranaree University of Technology (SUT) and Office of the Higher Education Commission under NRU project of Thailand, Thailand;
Turkish Atomic Energy Agency (TAEK), Turkey;
National Academy of  Sciences of Ukraine, Ukraine;
Science and Technology Facilities Council (STFC), United Kingdom;
National Science Foundation of the United States of America (NSF) and United States Department of Energy, Office of Nuclear Physics (DOE NP), United States of America.
\end{acknowledgement}

\bibliographystyle{utphys}   
\bibliography{Dmeson_v2}

\providecommand{\href}[2]{#2}\begingroup\raggedright\begin{thebibliography}{10}

\bibitem{Karsch:2006xs}
F.~Karsch, ``{Lattice simulations of the thermodynamics of strongly interacting
  elementary particles and the exploration of new phases of matter in
  relativistic heavy ion collisions},''
  \href{http://dx.doi.org/10.1088/1742-6596/46/1/017}{{\em J. Phys. Conf. Ser.}
  {\bfseries 46} (2006) 122},
\href{http://arxiv.org/abs/hep-lat/0608003}{{\ttfamily arXiv:hep-lat/0608003
  [hep-lat]}}.

\bibitem{Borsanyi:2010bp}
{\bfseries Wuppertal-Budapest} Collaboration, S.~Borsanyi, Z.~Fodor,
  C.~Hoelbling, S.~D. Katz, S.~Krieg, C.~Ratti, and K.~K. Szabo, ``{Is there
  still any $T_c$ mystery in lattice QCD? Results with physical masses in the
  continuum limit III},'' \href{http://dx.doi.org/10.1007/JHEP09(2010)073}{{\em
  JHEP} {\bfseries 09} (2010) 073},
\href{http://arxiv.org/abs/1005.3508}{{\ttfamily arXiv:1005.3508 [hep-lat]}}.

\bibitem{Bazavov:2011nk}
A.~Bazavov {\em et~al.}, ``{The chiral and deconfinement aspects of the QCD
  transition},'' \href{http://dx.doi.org/10.1103/PhysRevD.85.054503}{{\em Phys.
  Rev.} {\bfseries D85} (2012) 054503},
\href{http://arxiv.org/abs/1111.1710}{{\ttfamily arXiv:1111.1710 [hep-lat]}}.

\bibitem{FLORKOWSKI2016669}
W.~Florkowski, R.~Ryblewski, N.~Su, and K.~Tywoniuk, ``Strong-coupling effects
  in a plasma of confining gluons,''
  \href{http://dx.doi.org/http://dx.doi.org/10.1016/j.nuclphysa.2016.01.063}{{\em
  Nuclear Physics A} {\bfseries 956} (2016) 669}.
  \url{http://www.sciencedirect.com/science/article/pii/S0375947416000774}.

\bibitem{PhysRevD.85.054503}
{\bfseries HotQCD Collaboration} Collaboration, A.~Bazavov, T.~Bhattacharya,
  M.~Cheng, C.~DeTar, H.-T. Ding, S.~Gottlieb, R.~Gupta, P.~Hegde, U.~M.
  Heller, F.~Karsch, E.~Laermann, L.~Levkova, S.~Mukherjee, P.~Petreczky,
  C.~Schmidt, R.~A. Soltz, W.~Soeldner, R.~Sugar, D.~Toussaint, W.~Unger, and
  P.~Vranas, ``Chiral and deconfinement aspects of the qcd transition,''
  \href{http://dx.doi.org/10.1103/PhysRevD.85.054503}{{\em Phys. Rev. D}
  {\bfseries 85} (2012) 054503}.

\bibitem{BraunMunzinger:2007tn}
P.~Braun-Munzinger, ``{Quarkonium production in ultra-relativistic nuclear
  collisions: Suppression versus enhancement},''
  \href{http://dx.doi.org/10.1088/0954-3899/34/8/S36}{{\em J. Phys.} {\bfseries
  G34} (2007) S471},
\href{http://arxiv.org/abs/nucl-th/0701093}{{\ttfamily arXiv:nucl-th/0701093
  [NUCL-TH]}}.

\bibitem{Andronic:2015wma}
A.~Andronic {\em et~al.}, ``{Heavy-flavour and quarkonium production in the LHC
  era: from proton–-proton to heavy-ion collisions},''
  \href{http://dx.doi.org/10.1140/epjc/s10052-015-3819-5}{{\em Eur. Phys. J.}
  {\bfseries C76} (2016) 107},
\href{http://arxiv.org/abs/1506.03981}{{\ttfamily arXiv:1506.03981 [nucl-ex]}}.

\bibitem{Braaten:1991we}
E.~Braaten and M.~H. Thoma, ``{Energy loss of a heavy quark in the quark -
  gluon plasma},''
\href{http://dx.doi.org/10.1103/PhysRevD.44.R2625}{{\em Phys. Rev.} {\bfseries
  D44} (1991) R2625}.

\bibitem{Gyulassy:1990ye}
M.~Gyulassy and M.~Plumer, ``{Jet Quenching in Dense Matter},''
\href{http://dx.doi.org/10.1016/0370-2693(90)91409-5}{{\em Phys. Lett.}
  {\bfseries B243} (1990) 432}.

\bibitem{Baier:1996sk}
R.~Baier, Y.~L. Dokshitzer, A.~H. Mueller, S.~Peigne, and D.~Schiff,
  ``{Radiative energy loss and p(T) broadening of high-energy partons in
  nuclei},'' \href{http://dx.doi.org/10.1016/S0550-3213(96)00581-0}{{\em Nucl.
  Phys.} {\bfseries B484} (1997) 265},
\href{http://arxiv.org/abs/hep-ph/9608322}{{\ttfamily arXiv:hep-ph/9608322
  [hep-ph]}}.

\bibitem{Prino:2016cni}
F.~Prino and R.~Rapp, ``{Open Heavy Flavor in QCD Matter and in Nuclear
  Collisions},'' \href{http://dx.doi.org/10.1088/0954-3899/43/9/093002}{{\em J.
  Phys.} {\bfseries G43} (2016) 093002},
\href{http://arxiv.org/abs/1603.00529}{{\ttfamily arXiv:1603.00529 [nucl-ex]}}.

\bibitem{Adare:2010de}
{\bfseries PHENIX} Collaboration, A.~Adare {\em et~al.}, ``{Heavy Quark
  Production in $p+p$ and Energy Loss and Flow of Heavy Quarks in Au+Au
  Collisions at $\sqrt{s_{\rm NN}}=200$ GeV},''
  \href{http://dx.doi.org/10.1103/PhysRevC.84.044905}{{\em Phys. Rev.}
  {\bfseries C84} (2011) 044905},
\href{http://arxiv.org/abs/1005.1627}{{\ttfamily arXiv:1005.1627 [nucl-ex]}}.

\bibitem{Abelev:2006db}
{\bfseries STAR} Collaboration, B.~I. Abelev {\em et~al.}, ``{Transverse
  momentum and centrality dependence of high-$p_T$ non-photonic electron
  suppression in Au+Au collisions at $\sqrt{s_{\rm NN}} = 200$\,GeV},''
  \href{http://dx.doi.org/10.1103/PhysRevLett.106.159902}{{\em Phys. Rev.
  Lett.} {\bfseries 98} (2007) 192301},
  \href{http://arxiv.org/abs/nucl-ex/0607012}{{\ttfamily arXiv:nucl-ex/0607012
  [nucl-ex]}}.
[Erratum: Phys. Rev. Lett.106 (2011) 159902].

\bibitem{Adamczyk:2014uip}
{\bfseries STAR} Collaboration, L.~Adamczyk {\em et~al.}, ``{Observation of
  $D^0$ Meson Nuclear Modifications in Au+Au Collisions at $\sqrt{s_{\rm
  NN}}=200$ GeV},''
  \href{http://dx.doi.org/10.1103/PhysRevLett.113.142301}{{\em Phys. Rev.
  Lett.} {\bfseries 113} (2014) 142301},
\href{http://arxiv.org/abs/1404.6185}{{\ttfamily arXiv:1404.6185 [nucl-ex]}}.

\bibitem{Adler:2005xv}
{\bfseries PHENIX} Collaboration, S.~S. Adler {\em et~al.}, ``{Nuclear
  modification of electron spectra and implications for heavy quark energy loss
  in Au+Au collisions at $\sqrt{s_{\rm NN}} = 200$ GeV},''
  \href{http://dx.doi.org/10.1103/PhysRevLett.96.032301}{{\em Phys. Rev. Lett.}
  {\bfseries 96} (2006) 032301},
\href{http://arxiv.org/abs/nucl-ex/0510047}{{\ttfamily arXiv:nucl-ex/0510047
  [nucl-ex]}}.

\bibitem{Adam:2015sza}
{\bfseries ALICE} Collaboration, J.~Adam {\em et~al.}, ``{Transverse momentum
  dependence of D-meson production in Pb-Pb collisions at $
  \sqrt{s_{\mathrm{NN}}}=$ 2.76 TeV},''
  \href{http://dx.doi.org/10.1007/JHEP03(2016)081}{{\em JHEP} {\bfseries 03}
  (2016) 081},
\href{http://arxiv.org/abs/1509.06888}{{\ttfamily arXiv:1509.06888 [nucl-ex]}}.

\bibitem{Abelev:2012qh}
{\bfseries ALICE} Collaboration, B.~Abelev {\em et~al.}, ``{Production of muons
  from heavy flavour decays at forward rapidity in pp and Pb-Pb collisions at
  $\sqrt{s_{\rm NN}}$ = 2.76 TeV},''
  \href{http://dx.doi.org/10.1103/PhysRevLett.109.112301}{{\em Phys. Rev.
  Lett.} {\bfseries 109} (2012) 112301},
\href{http://arxiv.org/abs/1205.6443}{{\ttfamily arXiv:1205.6443 [hep-ex]}}.

\bibitem{Adam:2016khe}
{\bfseries ALICE} Collaboration, J.~Adam {\em et~al.}, ``{Measurement of the
  production of high-$p_{\rm T}$ electrons from heavy-flavour hadron decays in
  Pb-Pb collisions at $\mathbf{\sqrt{\it s_{\rm{NN}}}}$ = 2.76 TeV},''
\href{http://arxiv.org/abs/1609.07104}{{\ttfamily arXiv:1609.07104 [nucl-ex]}}.

\bibitem{Adam:2016wyz}
{\bfseries ALICE} Collaboration, J.~Adam {\em et~al.}, ``{Measurement of
  electrons from beauty-hadron decays in p-Pb collisions at
  $\mathbf{\sqrt{s_{\rm NN}}=5.02}$ TeV and Pb-Pb collisions at
  $\mathbf{\sqrt{s_{\rm NN}}=2.76}$ TeV},''
\href{http://arxiv.org/abs/1609.03898}{{\ttfamily arXiv:1609.03898 [nucl-ex]}}.

\bibitem{Khachatryan:2016ypw}
{\bfseries CMS} Collaboration, V.~Khachatryan {\em et~al.}, ``{Suppression and
  azimuthal anisotropy of prompt and nonprompt $J/\psi$ production in Pb-Pb
  collisions at $\sqrt{s_{\rm NN}}$ = 2.76 TeV},'' {\em Submitted to: Eur.
  Phys. J. C} (2016) ,
\href{http://arxiv.org/abs/1610.00613}{{\ttfamily arXiv:1610.00613 [nucl-ex]}}.

\bibitem{Qin:2010pf}
G.-Y. Qin, H.~Petersen, S.~A. Bass, and B.~Muller, ``{Translation of collision
  geometry fluctuations into momentum anisotropies in relativistic heavy-ion
  collisions},'' \href{http://dx.doi.org/10.1103/PhysRevC.82.064903}{{\em Phys.
  Rev.} {\bfseries C82} (2010) 064903},
\href{http://arxiv.org/abs/1009.1847}{{\ttfamily arXiv:1009.1847 [nucl-th]}}.

\bibitem{Voloshin:1994mz}
S.~Voloshin and Y.~Zhang, ``{Flow study in relativistic nuclear collisions by
  Fourier expansion of Azimuthal particle distributions},''
  \href{http://dx.doi.org/10.1007/s002880050141}{{\em Z. Phys.} {\bfseries C70}
  (1996) 665},
\href{http://arxiv.org/abs/hep-ph/9407282}{{\ttfamily arXiv:hep-ph/9407282
  [hep-ph]}}.

\bibitem{Poskanzer:1998yz}
A.~M. Poskanzer and S.~A. Voloshin, ``{Methods for analyzing anisotropic flow
  in relativistic nuclear collisions},''
  \href{http://dx.doi.org/10.1103/PhysRevC.58.1671}{{\em Phys. Rev.} {\bfseries
  C58} (1998) 1671},
\href{http://arxiv.org/abs/nucl-ex/9805001}{{\ttfamily arXiv:nucl-ex/9805001
  [nucl-ex]}}.

\bibitem{Ollitrault:1992bk}
J.-Y. Ollitrault, ``{Anisotropy as a signature of transverse collective
  flow},''
\href{http://dx.doi.org/10.1103/PhysRevD.46.229}{{\em Phys. Rev.} {\bfseries
  D46} (1992) 229}.

\bibitem{Batsouli:2002qf}
S.~Batsouli, S.~Kelly, M.~Gyulassy, and J.~L. Nagle, ``{Does the charm flow at
  RHIC?},'' \href{http://dx.doi.org/10.1016/S0370-2693(03)00175-8}{{\em Phys.
  Lett.} {\bfseries B557} (2003) 26},
\href{http://arxiv.org/abs/nucl-th/0212068}{{\ttfamily arXiv:nucl-th/0212068
  [nucl-th]}}.

\bibitem{Gyulassy:2000gk}
M.~Gyulassy, I.~Vitev, and X.~N. Wang, ``{High p(T) azimuthal asymmetry in
  noncentral A+A at RHIC},''
  \href{http://dx.doi.org/10.1103/PhysRevLett.86.2537}{{\em Phys. Rev. Lett.}
  {\bfseries 86} (2001) 2537},
\href{http://arxiv.org/abs/nucl-th/0012092}{{\ttfamily arXiv:nucl-th/0012092
  [nucl-th]}}.

\bibitem{Shuryak:2001me}
E.~V. Shuryak, ``{The Azimuthal asymmetry at large p(t) seem to be too large
  for a `jet quenching'},''
  \href{http://dx.doi.org/10.1103/PhysRevC.66.027902}{{\em Phys. Rev.}
  {\bfseries C66} (2002) 027902},
\href{http://arxiv.org/abs/nucl-th/0112042}{{\ttfamily arXiv:nucl-th/0112042
  [nucl-th]}}.

\bibitem{Molnar:2004ph}
D.~Molnar, ``{Charm elliptic flow from quark coalescence dynamics},''
  \href{http://dx.doi.org/10.1088/0954-3899/31/4/052}{{\em J. Phys.} {\bfseries
  G31} (2005) S421},
\href{http://arxiv.org/abs/nucl-th/0410041}{{\ttfamily arXiv:nucl-th/0410041
  [nucl-th]}}.

\bibitem{Andronic:2003zv}
A.~Andronic, P.~Braun-Munzinger, K.~Redlich, and J.~Stachel, ``{Statistical
  hadronization of charm in heavy ion collisions at SPS, RHIC and LHC},''
  \href{http://dx.doi.org/10.1016/j.physletb.2003.07.066}{{\em Phys. Lett.}
  {\bfseries B571} (2003) 36},
\href{http://arxiv.org/abs/nucl-th/0303036}{{\ttfamily arXiv:nucl-th/0303036
  [nucl-th]}}.

\bibitem{Greco:2003vf}
V.~Greco, C.~M. Ko, and R.~Rapp, ``{Quark coalescence for charmed mesons in
  ultrarelativistic heavy ion collisions},''
  \href{http://dx.doi.org/10.1016/j.physletb.2004.06.064}{{\em Phys. Lett.}
  {\bfseries B595} (2004) 202},
\href{http://arxiv.org/abs/nucl-th/0312100}{{\ttfamily arXiv:nucl-th/0312100
  [nucl-th]}}.

\bibitem{He:2012df}
M.~He, R.~J. Fries, and R.~Rapp, ``{${\rm D_s}$-Meson as Quantitative Probe of
  Diffusion and Hadronization in Nuclear Collisions},''
  \href{http://dx.doi.org/10.1103/PhysRevLett.110.112301}{{\em Phys. Rev.
  Lett.} {\bfseries 110} (2013) 112301},
\href{http://arxiv.org/abs/1204.4442}{{\ttfamily arXiv:1204.4442 [nucl-th]}}.

\bibitem{Adamczyk:2014yew}
{\bfseries STAR} Collaboration, L.~Adamczyk {\em et~al.}, ``{Elliptic flow of
  electrons from heavy-flavor hadron decays in Au+Au collisions at
  $\sqrt{s_{\rm NN}} = $ 200, 62.4, and 39 GeV},''
  \href{http://dx.doi.org/10.1103/PhysRevC.95.034907}{{\em Phys. Rev.}
  {\bfseries C95} (2017) 034907},
\href{http://arxiv.org/abs/1405.6348}{{\ttfamily arXiv:1405.6348 [hep-ex]}}.

\bibitem{Adamczyk:2017xur}
{\bfseries STAR} Collaboration, L.~Adamczyk {\em et~al.}, ``{Measurement of
  $D^0$ Azimuthal Anisotropy at Midrapidity in Au+Au Collisions at
  $\sqrt{s_{\rm NN}}=200$~GeV},''
  \href{http://dx.doi.org/10.1103/PhysRevLett.118.212301}{{\em Phys. Rev.
  Lett.} {\bfseries 118} no.~21, (2017) 212301},
\href{http://arxiv.org/abs/1701.06060}{{\ttfamily arXiv:1701.06060 [nucl-ex]}}.

\bibitem{Abelev:2013lca}
{\bfseries ALICE} Collaboration, B.~Abelev {\em et~al.}, ``{D meson elliptic
  flow in non-central Pb-Pb collisions at $\sqrt{s_{\rm NN}}$ = 2.76 TeV},''
  \href{http://dx.doi.org/10.1103/PhysRevLett.111.102301}{{\em Phys. Rev.
  Lett.} {\bfseries 111} (2013) 102301},
\href{http://arxiv.org/abs/1305.2707}{{\ttfamily arXiv:1305.2707 [nucl-ex]}}.

\bibitem{Abelev:2014ipa}
{\bfseries ALICE} Collaboration, B.~Abelev {\em et~al.}, ``{Azimuthal
  anisotropy of D meson production in Pb-Pb collisions at $\sqrt{s_{\rm NN}} =
  2.76$ TeV},'' \href{http://dx.doi.org/10.1103/PhysRevC.90.034904}{{\em Phys.
  Rev.} {\bfseries C90} (2014) 034904},
\href{http://arxiv.org/abs/1405.2001}{{\ttfamily arXiv:1405.2001 [nucl-ex]}}.

\bibitem{Adam:2016ssk}
{\bfseries ALICE} Collaboration, J.~Adam {\em et~al.}, ``{Elliptic flow of
  electrons from heavy-flavour hadron decays at mid-rapidity in Pb-Pb
  collisions at $ \sqrt{{s}_{\mathrm{NN}}}=2.76 $ TeV},''
  \href{http://dx.doi.org/10.1007/JHEP09(2016)028}{{\em JHEP} {\bfseries 09}
  (2016) 028},
\href{http://arxiv.org/abs/1606.00321}{{\ttfamily arXiv:1606.00321 [nucl-ex]}}.

\bibitem{Adam:2015pga}
{\bfseries ALICE} Collaboration, J.~Adam {\em et~al.}, ``{Elliptic flow of
  muons from heavy-flavour hadron decays at forward rapidity in Pb-Pb
  collisions at $\sqrt{s_{\rm NN}}= 2.76$ TeV},''
  \href{http://dx.doi.org/10.1016/j.physletb.2015.11.059}{{\em Phys. Lett.}
  {\bfseries B753} (2016) 41},
\href{http://arxiv.org/abs/1507.03134}{{\ttfamily arXiv:1507.03134 [nucl-ex]}}.

\bibitem{Uphoff:2012gb}
J.~Uphoff, O.~Fochler, Z.~Xu, and C.~Greiner, ``{Open Heavy Flavor in Pb+Pb
  Collisions at $\sqrt{s}=2.76$ TeV within a Transport Model},''
  \href{http://dx.doi.org/10.1016/j.physletb.2012.09.069}{{\em Phys. Lett.}
  {\bfseries B717} (2012) 430},
\href{http://arxiv.org/abs/1205.4945}{{\ttfamily arXiv:1205.4945 [hep-ph]}}.

\bibitem{He:2014cla}
M.~He, R.~J. Fries, and R.~Rapp, ``{Heavy Flavor at the Large Hadron Collider
  in a Strong Coupling Approach},''
  \href{http://dx.doi.org/10.1016/j.physletb.2014.05.050}{{\em Phys. Lett.}
  {\bfseries B735} (2014) 445},
\href{http://arxiv.org/abs/1401.3817}{{\ttfamily arXiv:1401.3817 [nucl-th]}}.

\bibitem{Monteno:2011gq}
M.~Monteno, W.~M. Alberico, A.~Beraudo, A.~De~Pace, A.~Molinari, M.~Nardi, and
  F.~Prino, ``{Heavy-flavor dynamics in nucleus-nucleus collisions: from RHIC
  to LHC},'' \href{http://dx.doi.org/10.1088/0954-3899/38/12/124144}{{\em J.
  Phys.} {\bfseries G38} (2011) 124144},
\href{http://arxiv.org/abs/1107.0256}{{\ttfamily arXiv:1107.0256 [hep-ph]}}.

\bibitem{Djordjevic:2015hra}
M.~Djordjevic and M.~Djordjevic, ``{Predictions of heavy-flavor suppression at
  5.1 TeV Pb + Pb collisions at the CERN Large Hadron Collider},''
  \href{http://dx.doi.org/10.1103/PhysRevC.92.024918}{{\em Phys. Rev.}
  {\bfseries C92} (2015) 024918},
\href{http://arxiv.org/abs/1505.04316}{{\ttfamily arXiv:1505.04316 [nucl-th]}}.

\bibitem{Cao:2013ita}
S.~Cao, G.-Y. Qin, and S.~A. Bass, ``{Heavy-quark dynamics and hadronization in
  ultrarelativistic heavy-ion collisions: Collisional versus radiative energy
  loss},'' \href{http://dx.doi.org/10.1103/PhysRevC.88.044907}{{\em Phys. Rev.}
  {\bfseries C88} (2013) 044907},
\href{http://arxiv.org/abs/1308.0617}{{\ttfamily arXiv:1308.0617 [nucl-th]}}.

\bibitem{Song:2015ykw}
T.~Song, H.~Berrehrah, D.~Cabrera, W.~Cassing, and E.~Bratkovskaya, ``{Charm
  production in Pb + Pb collisions at energies available at the CERN Large
  Hadron Collider},'' \href{http://dx.doi.org/10.1103/PhysRevC.93.034906}{{\em
  Phys. Rev.} {\bfseries C93} (2016) 034906},
\href{http://arxiv.org/abs/1512.00891}{{\ttfamily arXiv:1512.00891 [nucl-th]}}.

\bibitem{Nahrgang:2013xaa}
M.~Nahrgang, J.~Aichelin, P.~B. Gossiaux, and K.~Werner, ``{Influence of
  hadronic bound states above $T_c$ on heavy-quark observables in Pb + Pb
  collisions at at the CERN Large Hadron Collider},''
  \href{http://dx.doi.org/10.1103/PhysRevC.89.014905}{{\em Phys. Rev.}
  {\bfseries C89} (2014) 014905},
\href{http://arxiv.org/abs/1305.6544}{{\ttfamily arXiv:1305.6544 [hep-ph]}}.

\bibitem{Uphoff:2014hza}
J.~Uphoff, O.~Fochler, Z.~Xu, and C.~Greiner, ``{Elastic and radiative heavy
  quark interactions in ultra-relativistic heavy-ion collisions},''
  \href{http://dx.doi.org/10.1088/0954-3899/42/11/115106}{{\em J. Phys.}
  {\bfseries G42} (2015) 115106},
\href{http://arxiv.org/abs/1408.2964}{{\ttfamily arXiv:1408.2964 [hep-ph]}}.

\bibitem{Beraudo:2014boa}
A.~Beraudo, A.~De~Pace, M.~Monteno, M.~Nardi, and F.~Prino, ``{Heavy flavors in
  heavy-ion collisions: quenching, flow and correlations},''
  \href{http://dx.doi.org/10.1140/epjc/s10052-015-3336-6}{{\em Eur. Phys. J.}
  {\bfseries C75} (2015) 121},
\href{http://arxiv.org/abs/1410.6082}{{\ttfamily arXiv:1410.6082 [hep-ph]}}.

\bibitem{Cao:2017hhk}
S.~Cao, T.~Luo, G.-Y. Qin, and X.-N. Wang, ``{Heavy and light flavor jet
  quenching at RHIC and LHC energies},''
\href{http://arxiv.org/abs/1703.00822}{{\ttfamily arXiv:1703.00822 [nucl-th]}}.

\bibitem{Moore:2004tg}
G.~D. Moore and D.~Teaney, ``{How much do heavy quarks thermalize in a heavy
  ion collision?},'' \href{http://dx.doi.org/10.1103/PhysRevC.71.064904}{{\em
  Phys. Rev.} {\bfseries C71} (2005) 064904},
\href{http://arxiv.org/abs/hep-ph/0412346}{{\ttfamily arXiv:hep-ph/0412346
  [hep-ph]}}.

\bibitem{Aamodt:2008zz}
{\bfseries ALICE} Collaboration, K.~Aamodt {\em et~al.}, ``{The ALICE
  experiment at the CERN LHC},''
\href{http://dx.doi.org/10.1088/1748-0221/3/08/S08002}{{\em JINST} {\bfseries
  3} (2008) S08002}.

\bibitem{Abelev:2014ffa}
{\bfseries ALICE} Collaboration, B.~Abelev {\em et~al.}, ``{Performance of the
  ALICE Experiment at the CERN LHC},''
  \href{http://dx.doi.org/10.1142/S0217751X14300440}{{\em Int. J. Mod. Phys.}
  {\bfseries A29} (2014) 1430044},
\href{http://arxiv.org/abs/1402.4476}{{\ttfamily arXiv:1402.4476 [nucl-ex]}}.

\bibitem{Abelev:2013qoq}
{\bfseries ALICE} Collaboration, B.~Abelev {\em et~al.}, ``{Centrality
  determination of Pb-Pb collisions at $\sqrt{s_{NN}}$ = 2.76 TeV with
  ALICE},'' \href{http://dx.doi.org/10.1103/PhysRevC.88.044909}{{\em Phys.
  Rev.} {\bfseries C88} (2013) 044909},
\href{http://arxiv.org/abs/1301.4361}{{\ttfamily arXiv:1301.4361 [nucl-ex]}}.

\bibitem{Adam:2015ptt}
{\bfseries ALICE} Collaboration, J.~Adam {\em et~al.}, ``{Centrality dependence
  of the charged-particle multiplicity density at midrapidity in Pb-Pb
  collisions at $\sqrt{s_{\rm NN}}$ = 5.02 TeV},''
  \href{http://dx.doi.org/10.1103/PhysRevLett.116.222302}{{\em Phys. Rev.
  Lett.} {\bfseries 116} (2016) 222302},
\href{http://arxiv.org/abs/1512.06104}{{\ttfamily arXiv:1512.06104 [nucl-ex]}}.

\bibitem{Adam:2015jda}
{\bfseries ALICE} Collaboration, J.~Adam {\em et~al.}, ``{Measurement of
  D$_{s}^{+}$ production and nuclear modification factor in Pb-Pb collisions at
  $ \sqrt{s_{\mathrm{NN}}}=$ 2.76 TeV},''
  \href{http://dx.doi.org/10.1007/JHEP03(2016)082}{{\em JHEP} {\bfseries 03}
  (2016) 082},
\href{http://arxiv.org/abs/1509.07287}{{\ttfamily arXiv:1509.07287 [nucl-ex]}}.

\bibitem{Olive:2016xmw}
{\bfseries Particle Data Group} Collaboration, C.~Patrignani {\em et~al.},
  ``{Review of Particle Physics},''
\href{http://dx.doi.org/10.1088/1674-1137/40/10/100001}{{\em Chin. Phys.}
  {\bfseries C40} (2016) 100001}.

\bibitem{PhysRevC.77.034904}
I.~Selyuzhenkov and S.~Voloshin, ``Effects of nonuniform acceptance in
  anisotropic flow measurements,''
  \href{http://dx.doi.org/10.1103/PhysRevC.77.034904}{{\em Phys. Rev. C}
  {\bfseries 77} (2008) 034904}.

\bibitem{Aichelin:2012ww}
J.~Aichelin, P.~B. Gossiaux, and T.~Gousset, ``{Radiative and Collisional
  Energy Loss of Heavy Quarks in Deconfined Matter},''
  \href{http://dx.doi.org/10.5506/APhysPolB.43.655}{{\em Acta Phys. Polon.}
  {\bfseries B43} (2012) 655},
\href{http://arxiv.org/abs/1201.4192}{{\ttfamily arXiv:1201.4192 [nucl-th]}}.

\bibitem{Greco:2007sz}
V.~Greco, H.~van Hees, and R.~Rapp, ``{Heavy-quark kinetics at RHIC and LHC},''
  in {\em {Nuclear physics. Proceedings, 23rd International Conference, INPC
  2007, Tokyo, Japan, June 3-8, 2007}}.
\newblock 2007.
\newblock
\href{http://arxiv.org/abs/0709.4452}{{\ttfamily arXiv:0709.4452 [hep-ph]}}.
\newblock

\bibitem{ALICE:2012ab}
{\bfseries ALICE} Collaboration, B.~Abelev {\em et~al.}, ``{Suppression of high
  transverse momentum D mesons in central Pb-Pb collisions at $\sqrt{s_{\rm
  NN}}=2.76$ TeV},'' \href{http://dx.doi.org/10.1007/JHEP09(2012)112}{{\em
  JHEP} {\bfseries 09} (2012) 112},
\href{http://arxiv.org/abs/1203.2160}{{\ttfamily arXiv:1203.2160 [nucl-ex]}}.

\bibitem{Cacciari:2012ny}
M.~Cacciari, S.~Frixione, N.~Houdeau, M.~L. Mangano, P.~Nason, {\em et~al.},
  ``{Theoretical predictions for charm and bottom production at the {LHC}},''
  \href{http://dx.doi.org/10.1007/JHEP10(2012)137}{{\em JHEP} {\bfseries 10}
  (2012) 137},
\href{http://arxiv.org/abs/1205.6344}{{\ttfamily arXiv:1205.6344 [hep-ph]}}.

\bibitem{Lange:2001uf}
D.~J. Lange, ``{The EvtGen particle decay simulation package},''
\href{http://dx.doi.org/10.1016/S0168-9002(01)00089-4}{{\em Nucl. Instrum.
  Meth.} {\bfseries A462} (2001) 152}.

\bibitem{Miller:2007ri}
M.~L. Miller, K.~Reygers, S.~J. Sanders, and P.~Steinberg, ``{Glauber modeling
  in high energy nuclear collisions},''
  \href{http://dx.doi.org/10.1146/annurev.nucl.57.090506.123020}{{\em Ann. Rev.
  Nucl. Part. Sci.} {\bfseries 57} (2007) 205},
\href{http://arxiv.org/abs/nucl-ex/0701025}{{\ttfamily arXiv:nucl-ex/0701025
  [nucl-ex]}}.

\bibitem{Adam:2015nna}
{\bfseries ALICE} Collaboration, J.~Adam {\em et~al.}, ``{Centrality dependence
  of high-$p_{\rm T}$ D meson suppression in Pb-Pb collisions at
  $\sqrt{s_{\mathrm{N}\mathrm{N}}}=2.76 $ TeV},''
  \href{http://dx.doi.org/10.1007/JHEP11(2015)205}{{\em JHEP} {\bfseries 11}
  (2015) 205},
\href{http://arxiv.org/abs/1506.06604}{{\ttfamily arXiv:1506.06604 [nucl-ex]}}.

\bibitem{Abelev:2012di}
{\bfseries ALICE} Collaboration, B.~Abelev {\em et~al.}, ``{Anisotropic flow of
  charged hadrons, pions and (anti-)protons measured at high transverse
  momentum in Pb-Pb collisions at $\sqrt{s_{\rm NN}}$=2.76 TeV},''
  \href{http://dx.doi.org/10.1016/j.physletb.2012.12.066}{{\em Phys. Lett.}
  {\bfseries B719} (2013) 18},
\href{http://arxiv.org/abs/1205.5761}{{\ttfamily arXiv:1205.5761 [nucl-ex]}}.

\bibitem{ALICE-PUBLIC-2015-003}
{\bfseries ALICE} Collaboration, ``{Supplemental figure: Anisotropic flow of
  charged hadrons, pions and (anti-)protons measured at high transverse
  momentum in Pb-Pb collisions at $\mathbf{\sqrt{{\textit s}_{\rm NN}}}$ = 2.76
  TeV},'' {\em Public note} (Aug, 2015) .
  \url{http://cds.cern.ch/record/2045885}.

\bibitem{Abelev:2014pua}
{\bfseries ALICE} Collaboration, B.~Abelev {\em et~al.}, ``{Elliptic flow of
  identified hadrons in Pb-Pb collisions at $ \sqrt{s_{\mathrm{NN}}}=2.76 $
  TeV},'' \href{http://dx.doi.org/10.1007/JHEP06(2015)190}{{\em JHEP}
  {\bfseries 06} (2015) 190},
\href{http://arxiv.org/abs/1405.4632}{{\ttfamily arXiv:1405.4632 [nucl-ex]}}.

\bibitem{Sirunyan:2017plt}
{\bfseries CMS} Collaboration, A.~M. Sirunyan {\em et~al.}, ``{Measurement of
  prompt D$^0$ meson azimuthal anisotropy in Pb--Pb collisions at $\sqrt{s_{\rm
  NN}}=5.02$~TeV},''
\href{http://arxiv.org/abs/1708.03497}{{\ttfamily arXiv:1708.03497 [nucl-ex]}}.

\bibitem{Adam:2016izf}
{\bfseries ALICE} Collaboration, J.~Adam {\em et~al.}, ``{Anisotropic flow of
  charged particles in Pb-Pb collisions at $\sqrt{s_{\rm NN}}=5.02$ TeV},''
  \href{http://dx.doi.org/10.1103/PhysRevLett.116.132302}{{\em Phys. Rev.
  Lett.} {\bfseries 116} (2016) 132302},
\href{http://arxiv.org/abs/1602.01119}{{\ttfamily arXiv:1602.01119 [nucl-ex]}}.

\bibitem{Xu:2017hgt}
Y.~Xu, M.~Nahrgang, J.~E. Bernhard, S.~Cao, and S.~A. Bass, ``{A data-driven
  analysis of the heavy quark transport coefficient},''
\href{http://arxiv.org/abs/1704.07800}{{\ttfamily arXiv:1704.07800 [nucl-th]}}.

\bibitem{Ding:2012sp}
H.~T. Ding, A.~Francis, O.~Kaczmarek, F.~Karsch, H.~Satz, and W.~Soeldner,
  ``{Charmonium properties in hot quenched lattice QCD},''
  \href{http://dx.doi.org/10.1103/PhysRevD.86.014509}{{\em Phys. Rev.}
  {\bfseries D86} (2012) 014509},
\href{http://arxiv.org/abs/1204.4945}{{\ttfamily arXiv:1204.4945 [hep-lat]}}.

\bibitem{Banerjee:2011ra}
D.~Banerjee, S.~Datta, R.~Gavai, and P.~Majumdar, ``{Heavy Quark Momentum
  Diffusion Coefficient from Lattice QCD},''
  \href{http://dx.doi.org/10.1103/PhysRevD.85.014510}{{\em Phys. Rev.}
  {\bfseries D85} (2012) 014510},
\href{http://arxiv.org/abs/1109.5738}{{\ttfamily arXiv:1109.5738 [hep-lat]}}.

\bibitem{Aamodt:2011mr}
{\bfseries ALICE} Collaboration, K.~Aamodt {\em et~al.}, ``{Two-pion
  Bose-Einstein correlations in central Pb-Pb collisions at $\sqrt{{s}_{\rm
  NN}} =$ 2.76 TeV},''
  \href{http://dx.doi.org/10.1016/j.physletb.2010.12.053}{{\em Phys. Lett.}
  {\bfseries B696} (2011) 328},
\href{http://arxiv.org/abs/1012.4035}{{\ttfamily arXiv:1012.4035 [nucl-ex]}}.

\end{thebibliography}\endgroup

\newpage
\appendix
\section{The ALICE Collaboration}
\label{app:collab}



\begingroup
\small
\begin{flushleft}
S.~Acharya$^\textrm{\scriptsize 139}$,
D.~Adamov\'{a}$^\textrm{\scriptsize 96}$,
J.~Adolfsson$^\textrm{\scriptsize 34}$,
M.M.~Aggarwal$^\textrm{\scriptsize 101}$,
G.~Aglieri Rinella$^\textrm{\scriptsize 35}$,
M.~Agnello$^\textrm{\scriptsize 31}$,
N.~Agrawal$^\textrm{\scriptsize 48}$,
Z.~Ahammed$^\textrm{\scriptsize 139}$,
N.~Ahmad$^\textrm{\scriptsize 17}$,
S.U.~Ahn$^\textrm{\scriptsize 80}$
% ...
\renewcommand\labelenumi{\textsuperscript{\theenumi}~}

\section*{Affiliation notes}
\renewcommand\theenumi{\roman{enumi}}
\begin{Authlist}
\item \Adef{0}Deceased
\item \Adef{idp1779984}{Also at: Dipartimento DET del Politecnico di Torino, Turin, Italy}
\item \Adef{idp1799376}{Also at: Georgia State University, Atlanta, Georgia, United States}
\item \Adef{idp4089904}{Also at: M.V. Lomonosov Moscow State University, D.V. Skobeltsyn Institute of Nuclear, Physics, Moscow, Russia}
\item \Adef{idp4447328}{Also at: Department of Applied Physics, Aligarh Muslim University, Aligarh, India}
\item \Adef{idp5423456}{Also at: Institute of Theoretical Physics, University of Wroclaw, Poland}
\end{Authlist}

\section*{Collaboration Institutes}
\renewcommand\theenumi{\arabic{enumi}~}

$^{1}$A.I. Alikhanyan National Science Laboratory (Yerevan Physics Institute) Foundation, Yerevan, Armenia
\\
$^{2}$Benem\'{e}rita Universidad Aut\'{o}noma de Puebla, Puebla, Mexico
\\
$^{3}$Bogolyubov Institute for Theoretical Physics, Kiev, Ukraine
\\
$^{4}$Bose Institute, Department of Physics 
and Centre for Astroparticle Physics and Space Science (CAPSS), Kolkata, India
\\
$^{5}$Budker Institute for Nuclear Physics, Novosibirsk, Russia
\\
$^{6}$California Polytechnic State University, San Luis Obispo, California, United States
\\
$^{7}$Central China Normal University, Wuhan, China
\\
$^{8}$Centre de Calcul de l'IN2P3, Villeurbanne, Lyon, France
\\
$^{9}$Centro de Aplicaciones Tecnol\'{o}gicas y Desarrollo Nuclear (CEADEN), Havana, Cuba
\\
$^{10}$Centro de Investigaciones Energ\'{e}ticas Medioambientales y Tecnol\'{o}gicas (CIEMAT), Madrid, Spain
\\
$^{11}$Centro de Investigaci\'{o}n y de Estudios Avanzados (CINVESTAV), Mexico City and M\'{e}rida, Mexico
\\
$^{12}$Centro Fermi - Museo Storico della Fisica e Centro Studi e Ricerche ``Enrico Fermi', Rome, Italy
\\
$^{13}$Chicago State University, Chicago, Illinois, United States
\\
$^{14}$China Institute of Atomic Energy, Beijing, China
\\
$^{15}$COMSATS Institute of Information Technology (CIIT), Islamabad, Pakistan
\\
$^{16}$Departamento de F\'{\i}sica de Part\'{\i}culas and IGFAE, Universidad de Santiago de Compostela, Santiago de Compostela, Spain
\\
$^{17}$Department of Physics, Aligarh Muslim University, Aligarh, India
\\
$^{18}$Department of Physics, Ohio State University, Columbus, Ohio, United States
\\
$^{19}$Department of Physics, Pusan National University, Pusan, Republic of Korea
\\
$^{20}$Department of Physics, Sejong University, Seoul, Republic of Korea
\\
$^{21}$Department of Physics, University of Oslo, Oslo, Norway
\\
$^{22}$Department of Physics and Technology, University of Bergen, Bergen, Norway
\\
$^{23}$Dipartimento di Fisica dell'Universit\`{a} 'La Sapienza'
and Sezione INFN, Rome, Italy
\\
$^{24}$Dipartimento di Fisica dell'Universit\`{a}
and Sezione INFN, Cagliari, Italy
\\
$^{25}$Dipartimento di Fisica dell'Universit\`{a}
and Sezione INFN, Trieste, Italy
\\
$^{26}$Dipartimento di Fisica dell'Universit\`{a}
and Sezione INFN, Turin, Italy
\\
$^{27}$Dipartimento di Fisica e Astronomia dell'Universit\`{a}
and Sezione INFN, Bologna, Italy
\\
$^{28}$Dipartimento di Fisica e Astronomia dell'Universit\`{a}
and Sezione INFN, Catania, Italy
\\
$^{29}$Dipartimento di Fisica e Astronomia dell'Universit\`{a}
and Sezione INFN, Padova, Italy
\\
$^{30}$Dipartimento di Fisica `E.R.~Caianiello' dell'Universit\`{a}
and Gruppo Collegato INFN, Salerno, Italy
\\
$^{31}$Dipartimento DISAT del Politecnico and Sezione INFN, Turin, Italy
\\
$^{32}$Dipartimento di Scienze e Innovazione Tecnologica dell'Universit\`{a} del Piemonte Orientale and INFN Sezione di Torino, Alessandria, Italy
\\
$^{33}$Dipartimento Interateneo di Fisica `M.~Merlin'
and Sezione INFN, Bari, Italy
\\
$^{34}$Division of Experimental High Energy Physics, University of Lund, Lund, Sweden
\\
$^{35}$European Organization for Nuclear Research (CERN), Geneva, Switzerland
\\
$^{36}$Excellence Cluster Universe, Technische Universit\"{a}t M\"{u}nchen, Munich, Germany
\\
$^{37}$Faculty of Engineering, Bergen University College, Bergen, Norway
\\
$^{38}$Faculty of Mathematics, Physics and Informatics, Comenius University, Bratislava, Slovakia
\\
$^{39}$Faculty of Nuclear Sciences and Physical Engineering, Czech Technical University in Prague, Prague, Czech Republic
\\
$^{40}$Faculty of Science, P.J.~\v{S}af\'{a}rik University, Ko\v{s}ice, Slovakia
\\
$^{41}$Faculty of Technology, Buskerud and Vestfold University College, Tonsberg, Norway
\\
$^{42}$Frankfurt Institute for Advanced Studies, Johann Wolfgang Goethe-Universit\"{a}t Frankfurt, Frankfurt, Germany
\\
$^{43}$Gangneung-Wonju National University, Gangneung, Republic of Korea
\\
$^{44}$Gauhati University, Department of Physics, Guwahati, India
\\
$^{45}$Helmholtz-Institut f\"{u}r Strahlen- und Kernphysik, Rheinische Friedrich-Wilhelms-Universit\"{a}t Bonn, Bonn, Germany
\\
$^{46}$Helsinki Institute of Physics (HIP), Helsinki, Finland
\\
$^{47}$Hiroshima University, Hiroshima, Japan
\\
$^{48}$Indian Institute of Technology Bombay (IIT), Mumbai, India
\\
$^{49}$Indian Institute of Technology Indore, Indore, India
\\
$^{50}$Indonesian Institute of Sciences, Jakarta, Indonesia
\\
$^{51}$INFN, Laboratori Nazionali di Frascati, Frascati, Italy
\\
$^{52}$INFN, Laboratori Nazionali di Legnaro, Legnaro, Italy
\\
$^{53}$INFN, Sezione di Bari, Bari, Italy
\\
$^{54}$INFN, Sezione di Bologna, Bologna, Italy
\\
$^{55}$INFN, Sezione di Cagliari, Cagliari, Italy
\\
$^{56}$INFN, Sezione di Catania, Catania, Italy
\\
$^{57}$INFN, Sezione di Padova, Padova, Italy
\\
$^{58}$INFN, Sezione di Roma, Rome, Italy
\\
$^{59}$INFN, Sezione di Torino, Turin, Italy
\\
$^{60}$INFN, Sezione di Trieste, Trieste, Italy
\\
$^{61}$Inha University, Incheon, Republic of Korea
\\
$^{62}$Institut de Physique Nucl\'eaire d'Orsay (IPNO), Universit\'e Paris-Sud, CNRS-IN2P3, Orsay, France
\\
$^{63}$Institute for Nuclear Research, Academy of Sciences, Moscow, Russia
\\
$^{64}$Institute for Subatomic Physics of Utrecht University, Utrecht, Netherlands
\\
$^{65}$Institute for Theoretical and Experimental Physics, Moscow, Russia
\\
$^{66}$Institute of Experimental Physics, Slovak Academy of Sciences, Ko\v{s}ice, Slovakia
\\
$^{67}$Institute of Physics, Academy of Sciences of the Czech Republic, Prague, Czech Republic
\\
$^{68}$Institute of Physics, Bhubaneswar, India
\\
$^{69}$Institute of Space Science (ISS), Bucharest, Romania
\\
$^{70}$Institut f\"{u}r Informatik, Johann Wolfgang Goethe-Universit\"{a}t Frankfurt, Frankfurt, Germany
\\
$^{71}$Institut f\"{u}r Kernphysik, Johann Wolfgang Goethe-Universit\"{a}t Frankfurt, Frankfurt, Germany
\\
$^{72}$Institut f\"{u}r Kernphysik, Westf\"{a}lische Wilhelms-Universit\"{a}t M\"{u}nster, M\"{u}nster, Germany
\\
$^{73}$Instituto de Ciencias Nucleares, Universidad Nacional Aut\'{o}noma de M\'{e}xico, Mexico City, Mexico
\\
$^{74}$Instituto de F\'{i}sica, Universidade Federal do Rio Grande do Sul (UFRGS), Porto Alegre, Brazil
\\
$^{75}$Instituto de F\'{\i}sica, Universidad Nacional Aut\'{o}noma de M\'{e}xico, Mexico City, Mexico
\\
$^{76}$IRFU, CEA, Universit\'{e} Paris-Saclay, Saclay, France
\\
$^{77}$iThemba LABS, National Research Foundation, Somerset West, South Africa
\\
$^{78}$Joint Institute for Nuclear Research (JINR), Dubna, Russia
\\
$^{79}$Konkuk University, Seoul, Republic of Korea
\\
$^{80}$Korea Institute of Science and Technology Information, Daejeon, Republic of Korea
\\
$^{81}$KTO Karatay University, Konya, Turkey
\\
$^{82}$Laboratoire de Physique Corpusculaire (LPC), Clermont Universit\'{e}, Universit\'{e} Blaise Pascal, CNRS--IN2P3, Clermont-Ferrand, France
\\
$^{83}$Laboratoire de Physique Subatomique et de Cosmologie, Universit\'{e} Grenoble-Alpes, CNRS-IN2P3, Grenoble, France
\\
$^{84}$Lawrence Berkeley National Laboratory, Berkeley, California, United States
\\
$^{85}$Moscow Engineering Physics Institute, Moscow, Russia
\\
$^{86}$Nagasaki Institute of Applied Science, Nagasaki, Japan
\\
$^{87}$National and Kapodistrian University of Athens, Physics Department, Athens, Greece
\\
$^{88}$National Centre for Nuclear Studies, Warsaw, Poland
\\
$^{89}$National Institute for Physics and Nuclear Engineering, Bucharest, Romania
\\
$^{90}$National Institute of Science Education and Research, HBNI, Jatni, India
\\
$^{91}$National Nuclear Research Center, Baku, Azerbaijan
\\
$^{92}$National Research Centre Kurchatov Institute, Moscow, Russia
\\
$^{93}$Niels Bohr Institute, University of Copenhagen, Copenhagen, Denmark
\\
$^{94}$Nikhef, Nationaal instituut voor subatomaire fysica, Amsterdam, Netherlands
\\
$^{95}$Nuclear Physics Group, STFC Daresbury Laboratory, Daresbury, United Kingdom
\\
$^{96}$Nuclear Physics Institute, Academy of Sciences of the Czech Republic, \v{R}e\v{z} u Prahy, Czech Republic
\\
$^{97}$Oak Ridge National Laboratory, Oak Ridge, Tennessee, United States
\\
$^{98}$Petersburg Nuclear Physics Institute, Gatchina, Russia
\\
$^{99}$Physics Department, Creighton University, Omaha, Nebraska, United States
\\
$^{100}$Physics department, Faculty of science, University of Zagreb, Zagreb, Croatia
\\
$^{101}$Physics Department, Panjab University, Chandigarh, India
\\
$^{102}$Physics Department, University of Cape Town, Cape Town, South Africa
\\
$^{103}$Physics Department, University of Jammu, Jammu, India
\\
$^{104}$Physics Department, University of Rajasthan, Jaipur, India
\\
$^{105}$Physikalisches Institut, Eberhard Karls Universit\"{a}t T\"{u}bingen, T\"{u}bingen, Germany
\\
$^{106}$Physikalisches Institut, Ruprecht-Karls-Universit\"{a}t Heidelberg, Heidelberg, Germany
\\
$^{107}$Physik Department, Technische Universit\"{a}t M\"{u}nchen, Munich, Germany
\\
$^{108}$Purdue University, West Lafayette, Indiana, United States
\\
$^{109}$Research Division and ExtreMe Matter Institute EMMI, GSI Helmholtzzentrum f\"ur Schwerionenforschung GmbH, Darmstadt, Germany
\\
$^{110}$Rudjer Bo\v{s}kovi\'{c} Institute, Zagreb, Croatia
\\
$^{111}$Russian Federal Nuclear Center (VNIIEF), Sarov, Russia
\\
$^{112}$Saha Institute of Nuclear Physics, Kolkata, India
\\
$^{113}$School of Physics and Astronomy, University of Birmingham, Birmingham, United Kingdom
\\
$^{114}$Secci\'{o}n F\'{\i}sica, Departamento de Ciencias, Pontificia Universidad Cat\'{o}lica del Per\'{u}, Lima, Peru
\\
$^{115}$SSC IHEP of NRC Kurchatov institute, Protvino, Russia
\\
$^{116}$Stefan Meyer Institut f\"{u}r Subatomare Physik (SMI), Vienna, Austria
\\
$^{117}$SUBATECH, IMT Atlantique, Universit\'{e} de Nantes, CNRS-IN2P3, Nantes, France
\\
$^{118}$Suranaree University of Technology, Nakhon Ratchasima, Thailand
\\
$^{119}$Technical University of Ko\v{s}ice, Ko\v{s}ice, Slovakia
\\
$^{120}$Technical University of Split FESB, Split, Croatia
\\
$^{121}$The Henryk Niewodniczanski Institute of Nuclear Physics, Polish Academy of Sciences, Cracow, Poland
\\
$^{122}$The University of Texas at Austin, Physics Department, Austin, Texas, United States
\\
$^{123}$Universidad Aut\'{o}noma de Sinaloa, Culiac\'{a}n, Mexico
\\
$^{124}$Universidade de S\~{a}o Paulo (USP), S\~{a}o Paulo, Brazil
\\
$^{125}$Universidade Estadual de Campinas (UNICAMP), Campinas, Brazil
\\
$^{126}$Universidade Federal do ABC, Santo Andre, Brazil
\\
$^{127}$University of Houston, Houston, Texas, United States
\\
$^{128}$University of Jyv\"{a}skyl\"{a}, Jyv\"{a}skyl\"{a}, Finland
\\
$^{129}$University of Liverpool, Liverpool, United Kingdom
\\
$^{130}$University of Tennessee, Knoxville, Tennessee, United States
\\
$^{131}$University of the Witwatersrand, Johannesburg, South Africa
\\
$^{132}$University of Tokyo, Tokyo, Japan
\\
$^{133}$University of Tsukuba, Tsukuba, Japan
\\
$^{134}$Universit\'{e} de Lyon, Universit\'{e} Lyon 1, CNRS/IN2P3, IPN-Lyon, Villeurbanne, Lyon, France
\\
$^{135}$Universit\'{e} de Strasbourg, CNRS, IPHC UMR 7178, F-67000 Strasbourg, France, Strasbourg, France
\\
$^{136}$Universit\`{a} degli Studi di Pavia, Pavia, Italy
\\
$^{137}$Universit\`{a} di Brescia, Brescia, Italy
\\
$^{138}$V.~Fock Institute for Physics, St. Petersburg State University, St. Petersburg, Russia
\\
$^{139}$Variable Energy Cyclotron Centre, Kolkata, India
\\
$^{140}$Warsaw University of Technology, Warsaw, Poland
\\
$^{141}$Wayne State University, Detroit, Michigan, United States
\\
$^{142}$Wigner Research Centre for Physics, Hungarian Academy of Sciences, Budapest, Hungary
\\
$^{143}$Yale University, New Haven, Connecticut, United States
\\
$^{144}$Yonsei University, Seoul, Republic of Korea
\\
$^{145}$Zentrum f\"{u}r Technologietransfer und Telekommunikation (ZTT), Fachhochschule Worms, Worms, Germany
\endgroup

\end{document}